\documentclass[12pt]{article}
\usepackage{graphicx}
\usepackage{amssymb,amsmath,palatino,amsfonts,amsthm}
\usepackage{amssymb}
\usepackage{epstopdf}
\usepackage{slashed} 
\newcommand{\f}{\begin{equation}}
\newcommand{\ff}{\end{equation}}

\newcommand{\bea}{\begin{eqnarray}}
\newcommand{\eea}{\end{eqnarray}}

\begin{document}

%%%%%%%%%%%%%%%%%%%%%%%%%%%%%%%%%%%%%%%%%%%%%%%%
\title{The quantum  cosmological constant}
\author{Stephon Alexander, Joao Magueijo and 
Lee Smolin\thanks{lsmolin@perimeterinstitute.ca}
\\ \\
Department of Physics, Brown University, Providence, RI, 02906
\\
Theoretical Physics Group, The Blackett Laboratory, Imperial College,\\
 Prince Consort Rd., London, SW7 2BZ, United Kingdom
\\
Perimeter Institute for Theoretical Physics,\\
31 Caroline Street North, Waterloo, Ontario N2J 2Y5, Canada}
\date{\today}
\maketitle
%\vfill
\begin{abstract}

We present an extension of general relativity in which the cosmological constant  
becomes dynamical and turns out to be conjugate to the Chern-Simons invariant of the Ashtekar connection on a spatial slicing.  The latter has been proposed in \cite{Chopin-Lee} 
as a time variable for quantum gravity: the Chern-Simons time.  In the
quantum theory the inverse cosmological constant and Chern-Simons time will then become conjugate operators.  The "Kodama state" gets a new interpretation as a family of transition functions. 
These results imply an uncertainty relation between $\Lambda$ and Chern-Simons time;
the consequences of which will be discussed elsewhere.

%might be behind a solution to the cosmological constant problem, and is motivation for new models of the Universe. 
%
%We also connect the appearance of the Chern-Simons time with the chiral anomaly.  
%This reveals
%a new duality of the gravitational degrees of freedom,
%\f
%R^{AB} \rightarrow \frac{\Lambda}{3} \Sigma^{AB} \ \ \ \ \ \  
%\Sigma^{AB} \rightarrow \frac{3}{\Lambda} R^{AB} 
%\label{duality17}
%\ff
%where $\Sigma^{AB}$ is the self-dual two form of the metric and $R_{AB}$ is the left handed piece of the curvature tensor.

\end{abstract}
\newpage
\tableofcontents
%\newpage
\section{Introduction}

%\section{Making $\Lambda$ dynamical}

The long standing cosmological constant problem comes in many guises.  Before the observation of cosmic acceleration, the search had been to seek a theory where all contributions to the cosmological constant summed up to  zero.  If the data on acceleration continues to be consistent with a plain cosmological constant, rather than a more general form of dark energy, then the new cosmological constant problem becomes a question of extreme fine tuning.  In this work we take an agnostic position on this matter and are instead motivated by insights from non-perturbative quantum gravity.

In quantum mechanics, while perturbation theory is a powerful tool in unveiling numerous physical observables, it can often times obscure or be blind to important non-perturbative information.  For example, quantum tunnelling is a non-perturbative process and any perturbative expansion in the vicinity of the tunnelling barrier would be blind to tunnelling.  It is in this context we would like to address the issue of both the IR and UV versions of the cosmological constant problems\footnote{We will not address the issue of phase transitions in this work, and how it might impart on the problem.}.  In both cases, the evaluation of the vacuum energy arises from choosing a classical space time background and calculating the relevant perturbative vacuum diagrams, which are divergent and, after imposing regurlarization, deemed to be highly fine tuned.  

Given the difficulty of the problem,  perhaps a new approach is needed.  We explore the idea that the cosmological constant is a dynamical variable, not in the sense of a dynamical field, as in quintessence models, but as a degree of freedom for the entire spatial or spacetime manifold.  
Even though similar ideas have been explored before\cite{KP,testing,LS-GL}, the novelty of this paper is that it examines the issue from the
unique vantage point of the Ashtekar variables\cite{Abhay} 
and its associated quantization\cite{lqg-review}.  In this opening article we focus on the classical formulation of the theory.  This will give us guidance for turning the cosmological constant into an operator in the quantum theory, which we develop in later papers in this series.

In this work it is convenient to formulate general relativity by gauging the complexified Lorentz group, $SL(2,\mathbb{C})_\mathbb{C}$ on a four dimensional manifold $\cal M$. The space-time connection $A^{ab}=-A^{ba}$ is a one form, valued in $\mathfrak{sl}(2,\mathbb{C})_\mathbb{C}$ the Lie algebra of $SL(2,\mathbb{C})_\mathbb{C}$.  This Lie algebra is represented by complex antisymmetric, $4 \times 4$ matrices, $M^{ab}=-M^{ba}$, where $a,b,c=0,1,2,3$ are internal Lorentz indices.  The resulting gravitational dynamics is determined by a connection $A^{ab}$, a two form $\Sigma^{ab}$, valued in the Lie algebra of $SL(2,\mathbb{C})$ and a scalar field which provides a map $\Psi : \mathfrak{sl}(2,\mathbb{C})_C \rightarrow \mathfrak{sl}(2,\mathbb{C})_C$, which is written as
$\Psi_{abcd}$ with the following symmetries and constraints,
\f
\Psi_{abcd} =\Psi_{cdab} = - \Psi_{bacd}, \ \ \ \ \ \varepsilon^{abcd} \Psi_{abcd} =0
\ff
It is then convenient to change to two component spinor indices where $A,B=0,1$ are left handed spinor indices while $A',B'=0',1'$ are right handed spinor indices.   The connection then decomposes into
\f
A^{ab}=A^{AA' BB'} = \varepsilon^{AB} A^{A'B'} + A^{AB} \varepsilon^{A'B'}
\ff
and the two form $\Sigma^{ab}$ similarly decompose.   Very importantly, the curvature
two form decomposes the same way
\f
R^{ab}=R^{AA' BB'} = \varepsilon^{AB} R^{A'B'} + R^{AB} \varepsilon^{A'B'}
\ff
where
\f
R^{AB} = dA^{AB} + A^{AC} \wedge A_{C}^B
\ff
is the left handed part of the curvature tensor\footnote{In much of the literature on $LQG$, 
$R^{AB}$ is denoted $F^{AB}$.}

The scalar fields $\Psi_{abcd}$ decompose into pure spin two fields represented by
$\Psi_{ABCD}$ and $\Psi_{A'B'C'D'}$, both totally symmetric, and mixed components $\Psi_{ABA'B'}$ on symmetric pairs of indices.  Thus, 
\f
\Psi_{ABCD}= \Psi_{(ABCD)}
\ff
and the same for primed indices represents the spin two field.

In our work we formulate an extension of general relativity in which the cosmological constant, 
$\Lambda$, varies\footnote{Prior works in which $\Lambda$ varies have included
\cite{HT,KP,MarcH-uni,Rafael-uni,LS-GL}.}, and is determined by the solution of an equation of motion. The theory has contributions proportional to both $\Lambda$ and $\frac{1}{\Lambda}$ so the process that determines its value is non-perturbative in $\Lambda$.
% Similar to quantum tunnelling, we explore the idea that any perturbative evaluation of the cosmological constant is ill defined.  
The term in $\Lambda$ is of course just the spacetime volume:
\f
S^V = -\frac{1}{8 \pi G} \int_{\cal M} \frac{\Lambda}{6} \Sigma^{AB} \wedge \Sigma_{AB},
\label{SV}
\ff
where $ \Sigma^{AB}= e^{A'  A}  \wedge e^{B}_{A'}$ is the self-dual two form of the spacetime metric, $e^{B}_{A'}$. 
We propose a novel coupling
of $\frac{1}{\Lambda}$ to the topological invariant $\int_{\cal M} R^{AB} \wedge R_{AB}$,
where $R_{AB}$ is the left handed part of the curvature tensor.  
In the Euclidean theory which, for simplicity, we will be studying in this paper, this reads
\f
S^{new} = -\frac{1}{8 \pi G} \int_{\cal M} \frac{3}{2 \Lambda} R^{AB} \wedge R_{AB}.
\label{Snew}
\ff
This is motivated by a key
property of deSitter spacetime, which is that it is a self-dual solution, in the sense that
\f
R_{AB} = \frac{\Lambda}{3} \Sigma_{AB}.
\ff 
Thus, a formal duality
transformation that has deSitter spacetime as a fixed point, must have symmetry:
\f
R^{AB} \rightarrow \frac{\Lambda}{3} \Sigma^{AB} \ \ \ \ \ \  
\Sigma^{AB} \rightarrow \frac{3}{\Lambda} R^{AB} 
\label{duality0}
\ff
and therefore it must takes the usual cosmological constant term, (\ref{SV}), to its dual, which is $S^{new}$,
given by Eq.~(\ref{Snew}).

Right at the start we find an interesting quantum implication. 
A reflection of this duality is that the solution to the Hamilton-Jacobi equation that corresponds to deSitter spacetime is\cite{ls-positive} 
\f
S^{HJ}= \frac{3}{2 \Lambda} \int_\Sigma Y_{CS} (A)
\ff
where $ Y_{CS} (A)$ is the Chern-Simons invariant of the Ashtekar connection.
This is of course closely related to $S^{new}$, as  we will discuss below. It leads to 
to a semiclassical state,
\f
\Psi_K (A) = e^{\frac{\imath}{\hbar}  S^{HJ}}
\ff
which is called the Kodama state\cite{Kodama}.   Remarkably in some regularizations and ordering prescriptions this is an exact solution to all the constraints of quantum 
gravity\cite{ls-positive}. 
There have, however, been issues concerning the physical adequacy and interpretation of the Kodama state\cite{Witten-Kodama}.  Below we propose a new interpretation for  this state, stemming from our proposal. 

The form of the new term (\ref{Snew}), and particularly that it is $CP$ odd, suggests an anology between $\frac{3}{\Lambda}$ and the theta angle in $QCD$\cite{ABJ}.  Relating the cosmological constant problem to the theta vacuum was considered in the works of \cite{SA,AG1,AG2}.  This further suggests treating
$\Lambda$ as a dynamical degree of freedom.  We explore three versions of this idea, in which
$\Lambda$ is chosen to be a dynamical field, or a function of time in a preferred $3+1$ slicing, or a single variable for all spacetime\footnote{As in \cite{KP}.}.  
We discover in each case that the $\Lambda$ equation of motion determines its value.

Finally, in one case for the realization of this theory, we show how the Hamiltonian formalism might be set up and lead to the canonical quantization of the theory. Even ignoring details of the dynamics at the classical level,  we can see that a quantum uncertainty principle would always be in action, rendering Lambda and  Chern-Simons time\cite{Chopin-Lee}, 
complementary variables. This may possibly leads to deep implications for quantum cosmology and quantum gravity, which we outline in the concluding Section, and take up again elsewhere. 

Before beginning, we note that the idea that the cosmological constant is conjugate to a measure of time has appeared before, in the context of unimodular gravity\cite{MarcH-uni,Rafael-uni}.
There the conjugate measure of time is four volume to the past of a three slice.

%Here we obtain insight from a non-perturbative state of quantum gravity, the Kodama State as a transition function where the cosmological constant is an operator that is canonically conjugate to a natural time variable, the Chern-Simons time.  

%This gives us a precise prescription to obtain a modification of General relativity with a non-arbitrary and non-perturbative coupling of the cosmological to a Poyntrigain density, similar to an instanton number in Yang-Mills, with semblance to a CP violating theta parameter.  We obtain new equations of motion of the modified gravity theory which yields dynamics for the cosmological constant in three cases that is trivial in the absence of the Poyntrigin term.  We provide evidence that this modified theory, motivated from non-perturbative quantum gravity yields transitions that lead to the vanishing of the cosmological constant in a background independent manner.  This statement points to the perspective that quantum field theory in a curved background (background dependent) is the wrong starting point to evaluate the cosmological constant.

\section{A Plebanski formulation for our poposal}

We work first in the Euclidean case.  We fix a topology ${\cal M}= I \times \Sigma$,  where
$I$ is the interval and we take $\Sigma$ compact.
We start with an action for general relativity coupled to $N$ chiral fermion fields, all expressed in terms\footnote{Where $P_+^{abcd}$ is the projection operator onto self-dual two forms.}
 of Ashtekar variables\cite{Sam,Tedlee-action},
\begin{eqnarray}
\frac{1}{\hbar} S & =& \int_{\cal M}
\frac{1}{8 \pi G\hbar}  \left \{  - P_+^{abcd}  e_a \wedge e_b \wedge R_{cd}^{+} (A^+ )  
+ {2 \Lambda} \epsilon^{abcd}  e_a \wedge e_b \wedge e^c \wedge e^d
 +   \frac{3}{2 \Lambda}   R^{AB} \wedge R_{AB}  
\right \}
\nonumber \\
&&  +\Sigma_N \bar{\Psi}_{A^\prime} \sigma_a^{A^\prime B} e^a \wedge ({\cal D} \Psi )_B,  
\end{eqnarray}
but with the addition of a new, third term, which will suffice to make $\Lambda$ dynamical, as we shall presently see.
We divide the action by the Planck's constant for reasons to become apparent below. 

We then rewrite this action, with its new term, in the Plebanski formulation\cite{Plebanski,CDJ,Alexander:2012ge}:
\begin{eqnarray}
\frac{1}{\hbar} S^{Pleb} &=& \int_{\cal M} \frac{1}{8 \pi G \hbar} \left ( \Sigma^{AB} \wedge R_{AB} 
- \frac{\Lambda}{6} \Sigma^{AB} \wedge \Sigma_{AB}
- \frac{1}{2} \Phi_{ABCD} \Sigma^{AB} \wedge \Sigma^{CD}
\right.
\nonumber \\
&&- \left. \frac{3}{2 \Lambda}  R^{AB} \wedge R_{AB}  \right )
+ {\cal L}^{matter}
\label{Pleb}
\end{eqnarray}
%The new  term is
%\f
%S^{CS}= \frac{1}{16 \pi G \hbar} \int_{\cal M} \frac{3}{\Lambda}  R^{AB} \wedge R_{AB} 
%ff
The new term can be rewritten as:
\begin{eqnarray}
S^{CS}&=&- \frac{1}{16 \pi G \hbar}   \int_{\cal M}  \frac{3}{\Lambda}   R^{AB} \wedge R_{AB} 
=- \frac{1}{16 \pi G \hbar}  \int_{\cal M}  \frac{3}{\Lambda}  d Y_{CS} 
\\ \nonumber
&=& - \frac{1}{8 \pi G \hbar} \int_{\Sigma_{final}} \frac{3}{2 \Lambda} Y_{CS} + \frac{1}{8 \pi G \hbar}  \int_{\Sigma_{initial}} \frac{3}{2 \Lambda} Y_{CS} 
\nonumber \\
&+&  \frac{1}{16 \pi G \hbar}   \int_{\cal M} d (\frac{3}{\Lambda}) Y_{CS} 
\end{eqnarray}
We note that if we exponentiate this action (divided by $\hbar$), the first and second terms gives a generalization of 
the Kodama state on the initial and final slice. The last term can be written as:
\f
S^{CS}= \frac{3}{16 \pi \hbar G } \int dt \frac{\dot{\Lambda}}{\Lambda^2} \int_\Sigma Y_{CS}
\ff
and vanishes if Lambda is forced to be a constant.

The field equations for our theory, in the absence of matter, are:
\begin{eqnarray}
0= \frac{\delta S}{\delta \Phi_{ABCD}} & \rightarrow & \Sigma^{(AB} \wedge \Sigma^{CD)}=0
\label{SSe}
\\
0= \frac{\delta S}{\delta \Sigma_{AB}} & \rightarrow & R_{AB} = \frac{\Lambda}{3}  \Sigma_{AB}
+\Phi_{ABCD} \Sigma^{CD}
\label{Fem}
\\
0= \frac{\delta S}{\delta A_{AB}} & \rightarrow & 
{\cal D} \wedge \Sigma^{AB} \equiv  S^{AB} 
\label{torsionem}
\end{eqnarray}
with 
\f
S^{AB}= d(\frac{3}{ 2 \Lambda} )  \wedge R^{AB}=  
-\frac{3}{2 \Lambda^2} d\Lambda \wedge R^{AB}
\label{key1}
\ff
The solution to (\ref{SSe}) is
that there exists a frame field $e^{A A'}$ such that,
\f
\Sigma^{AB} = e^{A'  A}  \wedge e^{B}_{A'}
\label{frame}
\ff
is the self-dual two  form of the metric made from $e^{A A'}$. We also note that if
\f
 {\cal D} e^{B A'} =  T^{B A'} 
\ff
is the torsion, then
\f
S^{AB} = 2 e^{(A}_{\ A'} \wedge {\cal D} e^{B) A'} =
2 e^{(A}_{\ A'} \wedge T^{B)A'} 
\ff
A new feature is then a contribution to the torsion (\ref{key1})  related to the derivative of the cosmological constant.

\section{The underlying duality}

We can see that the terms that involve $\Lambda$:  
\f
S^\Lambda  = \frac{-1}{16 \pi \hbar G} \int_{\cal M} \left \{
\frac{\Lambda }{3}  \Sigma_{AB} \wedge \Sigma^{AB}
 + \frac{3}{ \Lambda}   R_{AB} \wedge R^{AB} \right \}
\ff
have an interesting structure:
%As announced in the Introduction, 
they have a formal duality symmetry under:
\f
R^{AB} \rightarrow \frac{\Lambda}{3} \Sigma^{AB} \ \ \ \ \ \  
\Sigma^{AB} \rightarrow \frac{3}{\Lambda} R^{AB}.
\label{duality}
\ff
The self-dual solutions, including deSitter spacetime, are the self-dual points:
\f
R^{AB} = \frac{\Lambda}{3} \Sigma^{AB} , \ \ \  \Phi_{ABCD}=0,
\label{sd}
\ff
where $\Phi_{ABCD}$ is a Lagrange multiplier. 

We say this symmetry is formal because on shell $\Sigma^{AB}$ satisfies
(\ref{SSe}), which is typically not satisfied by $R^{AB}$. 

Thus, we can extend the theory to one which has (\ref{duality}) as an exact symmetry:
\begin{eqnarray}
\frac{1}{\hbar} S^{Pleb} &=& \int_{\cal M} \frac{1}{8 \pi G \hbar} \left ( \Sigma^{AB} \wedge R_{AB} 
- \frac{\Lambda}{6} \Sigma^{AB} \wedge \Sigma_{AB}
- \frac{1}{2} \Phi_{ABCD} 
(\Sigma^{AB} + \frac{3}{\Lambda} R^{AB} ) \wedge ( \Sigma^{CD} +  \frac{3}{\Lambda} R^{CD} )
\right.
\nonumber \\
&&- \left. \frac{3}{2 \Lambda}  R^{AB} \wedge R_{AB}  \right )
+ {\cal L}^{matter}.
\end{eqnarray}
Instead of (\ref{frame}) this says that there is a frame field such that
\f
 e^{A'  A}  \wedge e^{B}_{A'} = \Sigma^{AB}  + \frac{3}{\Lambda} R^{AB} .
\ff
When we write the action and field equations in terms of this new $e^{A A'}$
we find this yields again an action for the Einstein equations.

\section{Three cases for the realization of the theory}\label{three}

The duality (\ref{duality}) results in the determination of $\Lambda$ as a function of the other fields.  To
see this we study the field equations for varying $\Lambda$.  There are
three cases, depending on what we choose $\Lambda$ to be a function of.

\begin{itemize} 

\item{}{\bf Case I: }$\Lambda (x^\mu ) $ is a function of space and time.

Varying by $\Lambda (x^\mu )$ we find an equation for $\Lambda (x^\mu )$:
\f
\frac{\Lambda (x^\mu )}{3} =   \sqrt{\frac{   R_{AB} \wedge R^{AB} }
{ \Sigma_{AB} \wedge \Sigma^{AB}    }}
\ff
Plugging this back into the action, we find
\f
S^\Lambda  = \frac{-1}{8 \pi \hbar G} \int_{\cal M}
 \sqrt{   R_{AB} \wedge R^{AB} }
 \sqrt{ \Sigma_{AB} \wedge \Sigma^{AB}}
\ff

This gives an interesting set of equations, the question is whether they are consistent.
Further study of this case is left to a future publication.

\item{}{\bf Case II: }$\Lambda$ is a function of time on some preferred $3+1$   slicing.

We fix an explicit slicing such as constant mean curvature slicing.  This gives a time
coordinate $t$. We also define Chern-Simons time by an integral over this slicing,
leading to $\tau_{CS} (t)$. 
%[JM: Imaginary part only]
We fix $\Lambda$ to be a function of the slicing.

Varying by $\Lambda (t)$ we find an equation for $\Lambda (t)$:
\f
\frac{\Lambda (t )}{3}=   \sqrt{\frac{\int_{\Sigma_t}   R_{AB} \wedge R^{AB}}
{\int_{\Sigma_t} \Sigma_{AB} \wedge \Sigma^{AB}    }}
\ff
Plugging this back into the action, we find
\f
S^\Lambda  = \frac{-1}{8 \pi \hbar G} \int dt
 \sqrt{\int_{\Sigma_t}   R_{AB} \wedge R^{AB} }
 \sqrt{\int_{\Sigma_t} \Sigma_{AB} \wedge \Sigma^{AB}}
\ff

This is similar to the theory described in \cite{LS-GL}, only rather than being conjugate to Newton's
constant, $G$, it appears the cosmological constant is conjugate to the Chern-Simons time $\tau_{CS}$
in the preferred slicing.

The equation of motion (\ref{Fem}) becomes non-local
\f
R_{AB} = 
+\Phi_{ABCD} \Sigma^{CD} +   \sqrt{\frac{\int_{\Sigma_t}   R_{AB} \wedge R^{AB}}
{\int_{\Sigma_t} \Sigma_{AB} \wedge \Sigma^{AB}    }} \Sigma_{AB}
\ff

This theory is also under investigation.  

%Indeed, the self-dual solutions, \ref{sd} satisfy this. 

We can check the homogeneous solutions, 
with\footnote{Note that the covariant  curl of (\ref{Fem}) vanishes because the torsion is (\ref{key1}). }
\f
R^{AB}= f(t) \Sigma^{AB}, \ \ \ \ \phi_{ABCD}=0
\ff
which yields
\f
f= \frac{\Lambda (t)}{3}
\ff
and the torsion (\ref{key1})
%Note that for the homogenous case without matter, this time dependence of $\Lambda$ is just 
%the trivial reparametrization invariance due to time dependence of the lapse\cite{ls-positive}.
%Is this right???
% CHECK AGAIN DF IS CONSISTEN IN Fem (13)  !!!!!!!!!!!

It is an important open question whether there are non-trivial solutions where $\Lambda (t)$
varies, with matter or non-zero Weyl tensor, which are not equivalent to deSitter spacetime.

\item{}{\bf Case III: }$\Lambda$ is one variable over all of spacetime\cite{KP}.

Varying by $\Lambda $ we find an equation for $\Lambda $:
\f
\frac{\Lambda }{3} =   \sqrt{\frac{\int_{\cal M}   R_{AB} \wedge R^{AB}}
{\int_{\cal M} \Sigma_{AB} \wedge \Sigma^{AB}    }}
%= \sqrt{12}   \sqrt{\frac{\int_{\Sigma_t}   dJ }{\int_{\Sigma_t} \Sigma_i \wedge \Sigma^i}}
\ff
Plugging this back into the action, we find
\f
S^\Lambda  = \frac{-1}{8 \pi \hbar G} 
 \sqrt{\int_{\cal M}   R_{AB} \wedge R^{AB} }
 \sqrt{\int_{\cal M} \Sigma_{AB} \wedge \Sigma^{AB}}
\ff

This is a version of the Kalapor-Padilla theory\cite{KP}

\end{itemize}

We can write out all the components of $R_{AB}$  as
\f
R_{AB}= \Psi_{ABCD} \Sigma^{CD} + R \Sigma_{AB} +\Phi_{AA' BB'} \bar{\Sigma}^{A'B'}
\ff
where $\Phi_{AA' BB'}$ and $R$ are, respectively, the trace-free part of the Ricci tensor
and its trace.  
Then, the Poyntriagin density is
\f
R^{AB} \wedge R_{AB} = 40 \imath e^4 \left (
R^2  + \Psi^{ABCD} \Psi_{ABCD}  -  \Phi^{AA' BB'} \Phi_{AA' BB'}
\right ).
\ff
It is interesting to note that in all cases the cosmological constant is proportional to the square root Pontryagin density.  It was shown that in metric variables
the Pontryagin density vanishes in most cosmological space-times.  However in approximate de-Sitter spacetimes, while the background Pontryagin density is zero, backreaceted chiral gravitational waves are sourced by a dynamical field that is coupled to the density and this will yield a non-vanishing Pontryagin density\cite{APS,AY}.  It is interesting to expect that such a mechanism could generate a small cosmological constant today and we leave this for a future work \cite{AJ}.

For the remainder of this paper we study Case II.

\section{The basis for a Hamiltonian treatment}\label{ham}
Let us now apply a $3+1$ decomposition to the action, which yields:
\f
S= \int dt \left [
\int_\Sigma \frac{1}{8 \pi G} \ (
E^{ai} \dot{A}_{ai} - {\cal N} {\cal H} -{\cal N}^a {\cal D}_a -\mu_i {\cal G}^i    )
+ \frac{3}{16 \pi \hbar G } \frac{ \dot{\Lambda}}{\Lambda^2} \int_\Sigma Y_{CS} (A)
\right ].
\ff
As explained above, 
we assume that $\Lambda$ is a function of time, $t$, alone. Then, we have, as usual, the  canonical brackets:
\f
\{ A_{a}^i (x) , E^b_k (y) \} = 8 \pi G \delta^b_a \delta^i_k \delta^3 (\tilde{y}, x ).
\ff
In addition  $\Lambda$ has a momentum $P$ such that,
\f
\{ \Lambda , P \} = 1
\label{[L,P]}
\ff
and there is a new primary constraint
% CHECK: FACTORS OF PI ETC
\f
{\cal W} = P-  \frac{3}{16 \pi  G }\frac{1 }{\Lambda^2} \int_\Sigma Y_{CS} (A) =0.
\label{W}
\ff
We add $\cal W$  to the Hamiltonian with a new lagrange multiplier, $\phi$.
\f
H=\int_\Sigma ( {\cal N} {\cal H}  + {\cal N}^a {\cal D}_a + \mu_i {\cal G}^i    )
+\phi {\cal W}.
\ff
In a future paper we will study further the algebra and other properties of this system of constraints. 
For the purpose of this paper it is sufficient to note that Lambda, or a function thereof, appears to be conjugate to a function 
of the Chern-Simons time\cite{Chopin-Lee} 
once the new primary constraint is taken into account. This suggests at once a quantum theory
containing a Heisenberg uncertainty principle involve Lambda and CS time, something we start to explore here. 

\section{Towards a quantum theory}
% CHECK FACTORS
Given the classical canonical structures unveiled in Section~\ref{ham} we can now lay down the basis of the quantum theory,
resulting in a new interpretation for the Kodama state. 
Combining (\ref{[L,P]}) and (\ref{W}) we can infer the Poisson bracket:
\f
\{ \Lambda , \int_\Sigma Y_{CS} (A) \} = \frac{16 \pi G \Lambda^2 }{3}.
\ff
Obtaining this classical structure was the ultimate goal of this first paper.
It suggests that in a quantum theory we can elevate $\Lambda$ and $\tau_{CS} = \int_\Sigma Y_{CS} (A)$
to operators with commutation relations,
\f\label{com1}
[ \hat{\Lambda}, \hat{\tau}_{CS} ] = \imath \frac{16 \pi \hbar G \hat{\Lambda}^2 }{3} 
\ff
We note that the commutator of $\Lambda$ is proportional to $\Lambda^2$,
so the larger $\Lambda$ is in Planck units the less classical it is.  Specifically, using purely
kinematical arguments, we can derive an uncertainty principle of the form:
\f
\Delta \Lambda \Delta \tau_{CS} \ge \frac{4 \pi \hbar G }{3} 
\langle \hat{\Lambda} \rangle^2.
\ff
If the expectation of Lambda is large in Planck units, then Lambda and CS time are complementary or incompatible variables. 
If it is not, as seems to be the case in the ``current''  Universe, they can be treated as classical variables. In our theory, 
the onset of classicality in cosmology is therefore related to the observed smallness
of Lambda.

Note that since $[q,p]=i$ implies $[f(q),p]=if'(q)$, we can re-express the commutator (\ref{com1}) 
in the more canonical form: 
\f
\left[ \hat{\frac{1}{\Lambda}}, \hat{\tau}_{CS} \right] = - \imath\frac{16 \pi \hbar G  }{3} .
\ff
It is then natural to consider representations that diagonalize either $\hat{\frac{1}{\Lambda}}$ or $\hat \tau_{CS}$, i.e. one of the two complementary variables. In the CS time representation we have:
\bea
\hat \tau_{CS}\Psi(\tau_{CS}) &=&\tau_{CS}\Psi(\tau_{CS}) \\
\widehat {\frac{1}{\Lambda}}\Psi(\tau_{CS}) &=&-i  \frac{4 \pi\hbar G}{3}\frac{\delta}{\delta \tau_{CS}}\Psi(\tau_{CS}).
\eea
The Kodama state then appears as an eigenstate of $\hat{\frac{1}{\Lambda}}$ in the CS time basis:
\f
\langle \tau_{CS}  |\frac{1}{\Lambda}  \rangle  = \Psi_{CS} ^\star = e^{-\imath \frac{3}{2\Lambda G \hbar}  \frac{1}{8\pi }
\tau_{CS} }= e^{-\imath \frac{3}{2\Lambda G \hbar}  \frac{1}{8\pi }\int_\Sigma Y_{CS}}.
\ff
More generally, the Kodama state can be seen as a transition amplitude between eigenstates of  $\hat \tau _{CS}$
and those of
$\hat{\frac{1}{\Lambda}}$:
\f
\langle \frac{1}{\Lambda} | \tau_{CS} \rangle = \Psi_{CS} = e^{\imath \frac{3}{2\Lambda G \hbar}  \frac{1}{8\pi }
\tau_{CS} }= e^{\imath \frac{3}{2\Lambda G \hbar}  \frac{1}{8\pi }\int_\Sigma Y_{CS}}.
\ff
Within a variable Lambda theory the Kodama state therefore receives a new 
interpretation as a transition amplitude.  
We further note that it satisfies a new operator self-dual equation with a quantum $\Lambda$ operator:
\f
\left ( \hat{E}^a_i -3  \epsilon^{abc}R_{bc}^i \hat{\Lambda}^{-1} \right  ) \Psi_{CS}=0.
\ff

%\section{Bringing in the anomaly.}

\section{Outlook}

In this paper we have introduced the term $S^{new}$, given by Eq.~(\ref{Snew}), aiming at introducing a 
variable or dynamical $\Lambda$,   and we have just begun a study of  the implications
for quantum gravity and cosmology.  At the level of the classical theory, we have left open
important questions.  To begin with we will want to extend these results 
to the Lorentzian theory, in which case the Chern-Simons time becomes the imaginary
part of the Chern-Simons invariant of the Ashtekar connection\cite{Chopin-Lee}.

In each of the three cases we have considered in Section~\ref{three} we need to establish whether
or not the field equations force $\Lambda$   to be constant.   If the field equations allow,
 we will want to study classical solutions where 
$\Lambda$ varies classically.  

We note that matter couplings may play a key role, as they may introduce terms in the torsion that compensate those due to derivatives of 
$\Lambda$, given by eq.  (\ref{key1}).
One reason to expect this will be noted below.

But even if $\Lambda$ is constrained to be constant at the classical level, there may be allowed
transitions in the quantum theory in which the value of $\Lambda$ changes.  These would be a new kind of tunnelling, which may be important in the early universe.  Naively, the amplitude
for such a transition would be proportional to
\f
{\cal A} = e^{\frac{\imath}{\hbar}  S^{\Lambda}}.
\ff
A quantum uncertainty principle between $\Lambda$  and a measure of
 time could also have deep cosmological consequences.  These will be the  subject
 of separate papers\cite{Q1}.
 
%If a structure like CS time is a relevant measure of cosmological time, which survives Quantum Gravity, and if this turns out to be incompatible with other observables essential for defining cosmological states, because there isa kinematic quantum uncertainty principle between it and $\Lambda$ , we may be forced into accepting a Universe delocalized in time, at least in some ``regions''. 

Further novelties in the the quantum theory would result from a possible
chiral gravitational anomaly. Let us write the quantum partition function as
\f
Z= \int De DA D\Lambda D\Psi D\bar{\Psi}    e^{\frac{\imath}{\hbar} S}.
\ff
We conjecture that under the integral $D\Psi D\bar{\Psi}$ for a set of chiral fermions, there will
be a chiral anomaly, 
\f
Tr F\wedge F = d J
\ff
where $J$ is the dual of the chiral current, $J_{abc}= \epsilon_{abcd} \tilde{J}^d$,
where
\f
\tilde{J}^a = 
\bar{\Psi}_{A^\prime} e^{A^\prime C a} \ \Psi_C
\label{chiral}
\ff
Note that in ordinary, perturbative quantum gravity, there is a chiral anomaly
of the form of (\ref{chiral}) so it is natural to conjecture that the anomaly appears
here as well. On the basis of this conjecture, we can write the action as
\begin{eqnarray}
S & =& \int_{\cal M} \left \{
\frac{1}{8 \pi G} (- \epsilon^{abcd}  e_a \wedge e_b \wedge R_{cd}^{+} (A^+ )  
+ {2 \Lambda} \epsilon^{abcd}  e_a \wedge e_b \wedge e^c \wedge e^d
 ) +\Sigma_N \bar{\Psi}_{A^\prime} \sigma_a^{A^\prime B} e^a \wedge ({\cal D} \Psi )_B
\right \} 
\nonumber \\
&& +\frac{3}{16 \pi G \hbar} \int_{\cal M}  \frac{1}{\Lambda} d J
\end{eqnarray}
Therefore, in our theory, we can write the CS phase as:
\f
S^{CS}=\frac{3}{16 \pi G \hbar} \int_{\cal M}   \frac{1}{\Lambda}  d J 
= -\frac{3}{16 \pi G \hbar} \int_{\cal M}   \frac{1}{\Lambda^2} d \Lambda  J
= +\frac{3}{16 \pi G \hbar} \int ds \frac{1}{\Lambda^2}\dot{\Lambda}  \int_{\Sigma} J
%=  \frac{3}{4 \pi \hbar G \Lambda^2 } \int ds \dot{\Lambda}  \int_{\Sigma}   J
\ff
We then have a torsion added to the connection
\f
T^a_i =  \frac{\delta S^{CS}}{\delta A_a^i} =  ( d(\frac{1}{\Lambda}) \wedge R^i )^{* a} = ( ( d(\frac{1}{\Lambda}) \wedge J^i )^{* a}
\ff
where
\f
\tilde{J}_i^a = \frac{\delta S^\Psi }{\delta A_a^i (x) } =  \frac{1}{4 \pi} 
\bar{\Psi}_{A^\prime} e^{A^\prime B a} \sigma_{i B }^C \Psi_C
\ff
This modifies the curvature tensor and the Einstein Equation. 
These aspects of the quantum theory will be explored in companion papers.

To conclude, we have laid down the basis for a new theory of a ``quantum cosmological constant'' that could address the
nagging problems $\Lambda$ leads to in cosmology and quantum gravity. The glimpses obtained already shed new light on outstanding 
issues, such as how to interpret the Kodama state in Quantum Gravity. Finally, as we will argue elsewhere, the quantum theory also hints at how a non-perturbative
approach might resolve the problem of the smallness of $\Lambda$ in our Universe. 

%We will also argue that a new model of the Universe, perhaps cyclic, could be build based on an idea of delocalized time, implicit in the complementary principle  we have unveiled. The philosophical implications of delocalized time are also not to be underestimated. 

\section*{Acknowledgements}

We are grateful to  Robert Brandenberger, Chris Hull, Evan McDonough, Laurent Freidel, 
Marc Henneaux, David Jennings, Sumati Surya,  Yigit Yarig, Francesca Vidotto and
Wolfgang Wieland for discussions 
and correspondence
 and Robert Sims for discussions and for comments on the manuscript.

This research was supported in part by Perimeter Institute for Theoretical Physics. Research at Perimeter Institute is supported by the Government of Canada through Industry Canada and by the Province of Ontario through the Ministry of Research and Innovation. This research was also partly supported by grants from NSERC and FQXi.  LS is especially thankful to the John Templeton Foundation for their generous support of this project. The work of JM was supported by a Consolidated STFC grant.

\end{document}